\documentclass[prb,aps,twocolumn,amsmath,amssymb,superscriptaddress]{revtex4}

\usepackage[dvips]{epsfig}
\usepackage{float}
\usepackage{tabularx}
\usepackage{makecell}
\usepackage{color}
\usepackage[sort&compress]{natbib}
\makeatletter
\renewcommand{\@biblabel}[1]{#1. }
\renewcommand{\@dotsep}{500}
\renewcommand{\@pnumwidth}{0em}
\renewcommand{\l@figure}[2]{
        \@dottedtocline{1}{1.5em}{2em}{Figure #1}{}\vspace{15pt}}

\begin{document}
\title{Silicon Photonic Entangled Photon-Pair and Heralded Single Photon Generation with {CAR} $>$ 12,000 and $g^{(2)}(0)<$ 0.006}
\author{Chaoxuan Ma}
\author{Xiaoxi Wang}
\address{University of California, San Diego, Department of Electrical and Computer Engineering, La~Jolla, California 92093-0407, USA}
\author{Vikas Anant}
\affiliation{Photon Spot Inc., 142 West Olive Avenue, Monrovia, California 91016, USA}
\author{Andrew D. Beyer}
\author{Matthew D. Shaw}
\affiliation{Jet Propulsion Laboratory, California Institute of Technology, 4800 Oak Grove Drive, Pasadena, California 91109, USA}
\author{Shayan Mookherjea}
\address{University of California, San Diego, Department of Electrical and Computer Engineering, La~Jolla, California 92093-0407, USA}
\email{smookherjea@ucsd.edu}
\date{\today}

\begin{abstract}
We report measurements of time-frequency entangled photon pairs and heralded single photons at 1550~nm wavelengths generated using a microring resonator pumped optically by a diode laser. Along with a high spectral brightness of pair generation, the conventional metrics used to describe performance, such as Coincidences-to-Accidentals Ratio (CAR), conditional self-correlation [$g^{(2)}(0)$], two-photon energy-time {F}ranson interferometric visibility etc. are shown to reach a high-performance regime not yet achieved by silicon photonics, and attained previously only by crystal, glass and fiber-based pair-generation devices.  
\end{abstract}

\maketitle
\section{Introduction}
There is recent interest in silicon photonic approaches to generating, manipulating and detecting quantum light, including entangled photon pairs and heralded single photons as resources for quantum optical communications and information processing. Integrated photonic structures, such as waveguides and micro-resonators, can be used for photon pair and heralded single photon generation using the nonlinear optical process of spontaneous four-wave mixing (SFWM)~\cite{Sharping2006,Clemmen2009,Azzini:12,davanco2012,PhysRevX.4.041047,Gentry15,Jiang:15,Savanier:16,Lu:16}. The intrinsic rate of nonlinear optical processes increases as the mode volume decreases, which reduces the pump power requirements of silicon microrings used for pair generation to sub-milliwatt levels\cite{Savanier:16}, and the device footprint to about one hundred square-microns\cite{Clemmen2009,Azzini:12}. Silicon device technology is useful not only for making high quality factor ($Q$) compact optical micro-resonators such as microrings and microdisks, but also driving and monitoring them with the help of micro-electronic components~\cite{savanier2015optimizing,Savanier2016a}. Also, the larger free-spectral range (FSR) compared to glass or silicon nitride micro-resonators (in the few-nanometer range) makes it easier to extract and measure the signal and idler photons from the ``comb-like'' multiplexed state where many wavelength-pairs are simultaneously generated\cite{Chen2011}, since the required components can be selected from commercially-available devices (such as filters, de-multiplexers, arrayed waveguide gratings, delay line interferometers etc.) which have already been designed for telecommunications.

However, the reported performance---in terms of the usual metrics such as Coincidences-to-Accidentals Ratio (CAR), conditional self-correlation [$g^{(2)}(0)$] and two-photon interference visibility (V)---has been significantly inferior to those of traditional pair-generation devices formed using optical fiber or crystals such as periodically-poled lithium niobate (PPLN) and potassium trihydrogen phosphate (KTP), where $\text{CAR} > 10,000$, $g^{2}(0) < 0.01$ and $V \simeq 99\%$ are common\cite{art-Brida-2012,Harder:13,Bock:16}. Silicon-photonic-based pair generation devices may not be able to generate a comparable number of photon pairs per second because of the weaker nonlinearity compared to crystal, and length limitations compared to fiber. But silicon volume manufacturing using wafer-scale technology can be inexpensive, and silicon photonic devices can be integrated with lasers\cite{Wang:17}, detectors\cite{schuck2013nbtin} and micro-electronics for future integrated systems. Thus, it is desirable at this time that a high \textit{quality\/} of the photon pairs be demonstrated. There are applications, such as detector calibration and short-range communications, where a very large number of photons per second is perhaps not essential, but a source that is neither bright nor of high quality is probably not of much use, even if cheap.  

The objective of this paper is to report record performance numbers achieved (at room-temperature, in an ``open'' setup, i.e., not sealed off from the laboratory environment) using optically-pumped SFWM in a high-Q silicon microring resonator. Previous reports have shown saturation / roll-off of parameters such as CAR at much lower values than measured here; therefore, these results present a more optimistic prospect for using silicon photonic devices for pair generation in quantum optics experiments than may have been expected so far.  
    
\section{Experimental details}
The microring was fabricated using a foundry silicon photonic process on SOI wafers, using ridge waveguides of width 0.65~$\mu\mathrm{m}$, height 0.22~$\mu\mathrm{m}$, and slab thickness 70~$\mathrm{nm}$, designed for low loss transmission in the lowest-order mode of the transverse electric (TE) polarization defined relative to the device plane. The microring had a radius $R=10\ \mu\mathrm{m}$. The slab regions of the ridge waveguides were doped, followed by contact and via formation and metalization, to form a p-i-n diode for monitoring, under reverse-bias, the optical power circulating in the microring. The Si waveguides used in the feeder waveguide and microring had a propagation loss (measured on test sites) of approximately 1 dB/cm, resulting in a microring intrinsic quality factor of approximately $9\times 10^5$, and a resonance lifetime $\tau \approx 76\ \mathrm{ps}$ (loaded quality factor of $9.2 \times 10^4$ at 1550~nm, with a spectral full-width at half-maximum (FWHM) of approximately 2.1 GHz).

\begin{figure*}[htb]
\includegraphics[width=5.8in, clip=true]{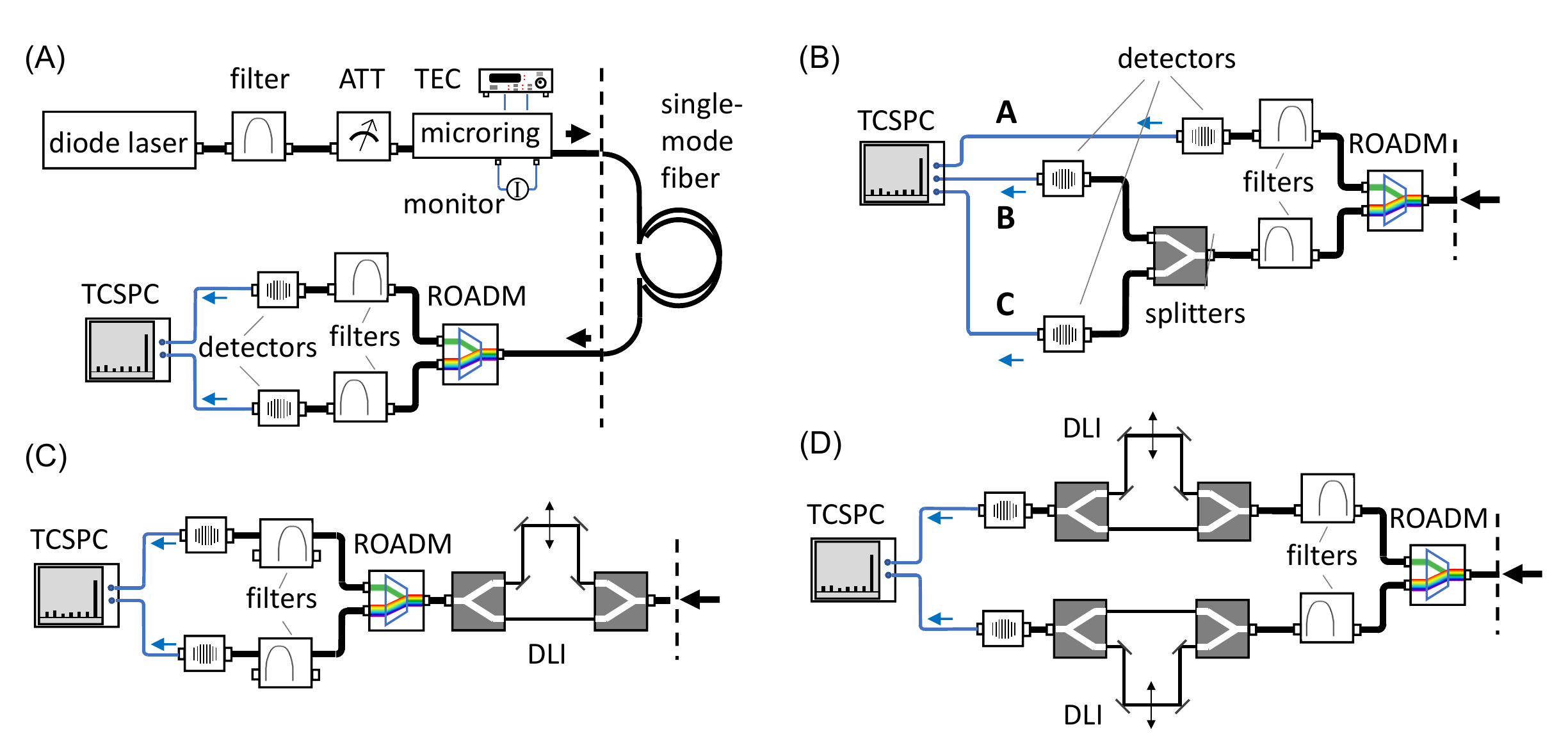}
\caption{(A) The experimental configuration for pair generation, and measurement of the coincidences-to-accidentals ratio (CAR). ATT: Variable optical attenuator, TEC: thermo-electric controller, TCSPC: time-correlated single-photon counter. (B) Modifications to the detection setup for measurement of the conditional (heralded) self-correlation, $g^{(2)}(0)$. (C) Measurement of two-photon Franson interferometric visibility using the folded configuration. DLI: Delay-Line Interferometer. (D) Measurement of two-photon Franson interferometric visibility using the un-folded configuration.} 
\label{fig-1}
\end{figure*}

Measurements reported here used the experimental configuration shown in Fig.~1. The bare-die chip was mounted on a temperature-controlled stage with a thermo-electric controller (TEC) in feedback with a thermistor on the stage mount. To establish a stable resonance, the silicon photonic chip with the microring was heated until the selected resonance aligned with the pump laser. The spectral alignment was continuously monitored during measurement using the reverse-biased photo-current of a silicon p-i-n junction diode fabricated across the microring\cite{savanier2015optimizing}, and confirmed using high-magnification infrared camera images of the microring. Polarization-maintaining fibers, fiber-loop paddles and lensed tapered fibers with anti-reflection coating were used to couple light to and from the silicon chip, and nanopositioning stages with piezoelectric actuators were used for accurate positioning of the fiber tips to the waveguide facets. An automated software program attempted to continuously optimize the coupling. The insertion loss of each fiber-to-waveguide coupler was estimated as 3.5~dB averaged over the wavelengths of interest based on previous experiments. Light was coupled to and from the chip using lensed tapered polarization-maintaining fibers. Output light from the chip was routed through cascaded filters to select one pair of spectral lines of Stokes (also called idler) and anti-Stokes (signal) photons positioned symmetrically around the pump wavelength. 

Under SFWM, energy-conservation between the pump and the generated Stokes and anti-Stokes photon pair dictates the frequency relationship, $2\omega_p = \omega_{S} + \omega_{aS}$, so that all three frequencies (wavelengths) lie within the band used in communication networks near 1550~nm. The microring provided simultaneous resonance for all three frequencies across adjacent free-spectral ranges with a tight constraint on the narrow bandwidth dictated by the high-quality resonance. The pump wavelength was positioned at 1554.9 nm and signal and idler photons were detected at 1535.5 nm and 1574.7 nm, respectively. External tunable filters (benchtop components) were used at these three wavelengths with FWHM's of approximately 1 nm, 0.6 nm and 0.8 nm, respectively. The spectral width of the microring resonance was approximately 0.03~nm, much narrower than the filter widths. Thus, the filters do not reshape the joint-spectral intensity, as may be a concern with broadband SFWM in waveguides.     

Photons were detected using fiber-coupled superconducting (WSi) nanowire single photon detectors (SNSPD), cooled to 0.8~K in a closed-cycle Helium-4 cryostat equipped with a sorption stage. The detection efficiencies for the SNSPDs were about 90\% for two detectors and about 65\% for the other two detectors; these detectors were not gated and were operated in a simple dc-biased mode with an RF-amplified readout. Coincidences were measured using a multi-input time-to-digital converter (TDC) instrument, with 0.08~ns minimum bin width, in start-stop mode. To prevent binning artifacts when accumulating histograms, at the cost of a factor-of-two in temporal resolution, two adjacent hardware bins were summed, according the manufacturer's suggestions, resulting in the 0.16~ns bin width used for all coincidence measurements. For confirmation, a few coincidence histograms were also measured using a two-channel  time-correlated single-photon counting (TCSPC) instrument with 25~ps bin size; these gave similar results in terms of the shape and widths of the coincidence peaks.

\section{Measurements}
Typical characterization measurements for SFWM report pair generation rates (PGR) and two-photon correlation and interference measurements, quantifying the pair generation, heralding and entanglement properties of the source\cite{Migdall2013}. Since the purpose of this paper is to demonstrate that high CAR values and low $g^{(2)}(0)$ values can be experimentally achieved using off-the-shelf, foundry-fabricated silicon photonic devices, we focus mainly on the low pump-power case; nevertheless, an appreciable rate of pairs and single photons was measured because of the relatively high brightness of the source. 
\begin{figure*}[ht]
\includegraphics[width=5.8in, clip=true]{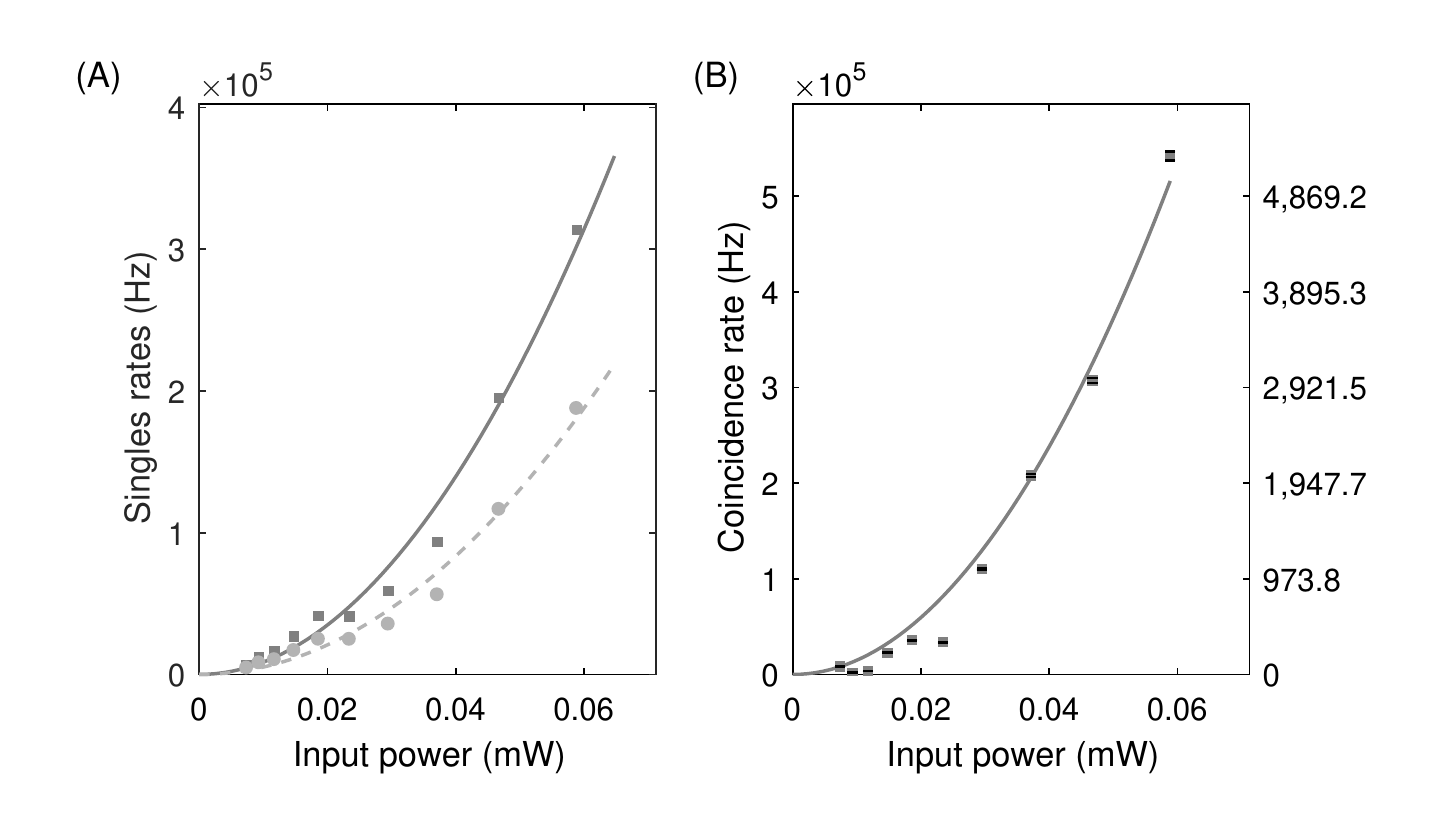}
\caption{(A) Singles count rates (Hz, raw measurements, not scaled) versus (cw) optical pump power in the feeder silicon waveguide. (B) Coincidence rate (in Hz), using the setup shown in Fig.~1(A). The left-hand side vertical axis shows the scaled coincidence count rates, accounting for chip coupling loss, filter insertion loss and detection efficiency. From the fit of this data, we infer the pair generation rate (PGR) as discussed in Section~\ref{sec-counts}. The right-hand side vertical axis shows the raw measured coincidence count rates (Hz).} 
\label{fig-2}
\end{figure*}

\subsection{Single and Coincidence Counts}
\label{sec-counts}
Fig.~2(A) shows the measured singles rates as a function of pump power, with differences in the values based on the slightly different losses through the filters (5.0 dB and 7.2 dB at the signal and idler wavelengths). Both sets of data are fitted by a quadratic function of the input pump power $P$ (in the feeder waveguide before the microring). The on-chip PGR was calculated from the measured coincidence rate by accounting for the insertion loss of the filters, and chip-waveguide coupling (3.5 dB), and the efficiency of the detectors (0.9). In Fig.~2(B), the (on-chip) pair generation rate is shown, and the fitted line, following the functional form $\mathrm{PGR} = R\times P^2$, agreed with the data. (The right-hand side axis shows the raw measured coincidence rates.) The fitted PGR is $R = 149\ \pm\ 6\ \mathrm{MHz}.\mathrm{mW}^{-2}$ (one standard deviation uncertainty). This is a good value for silicon microrings, which improves upon previous experimental results ($R = 1-10 \ \mathrm{MHz}.\mathrm{mW}^{-2}$) tabulated in Ref.~\cite{Savanier:16}, and leads to a high spectral brightness throughout the operating range, as described below. 

\begin{figure*}[ht]
\includegraphics[width=5.8in, clip=true]{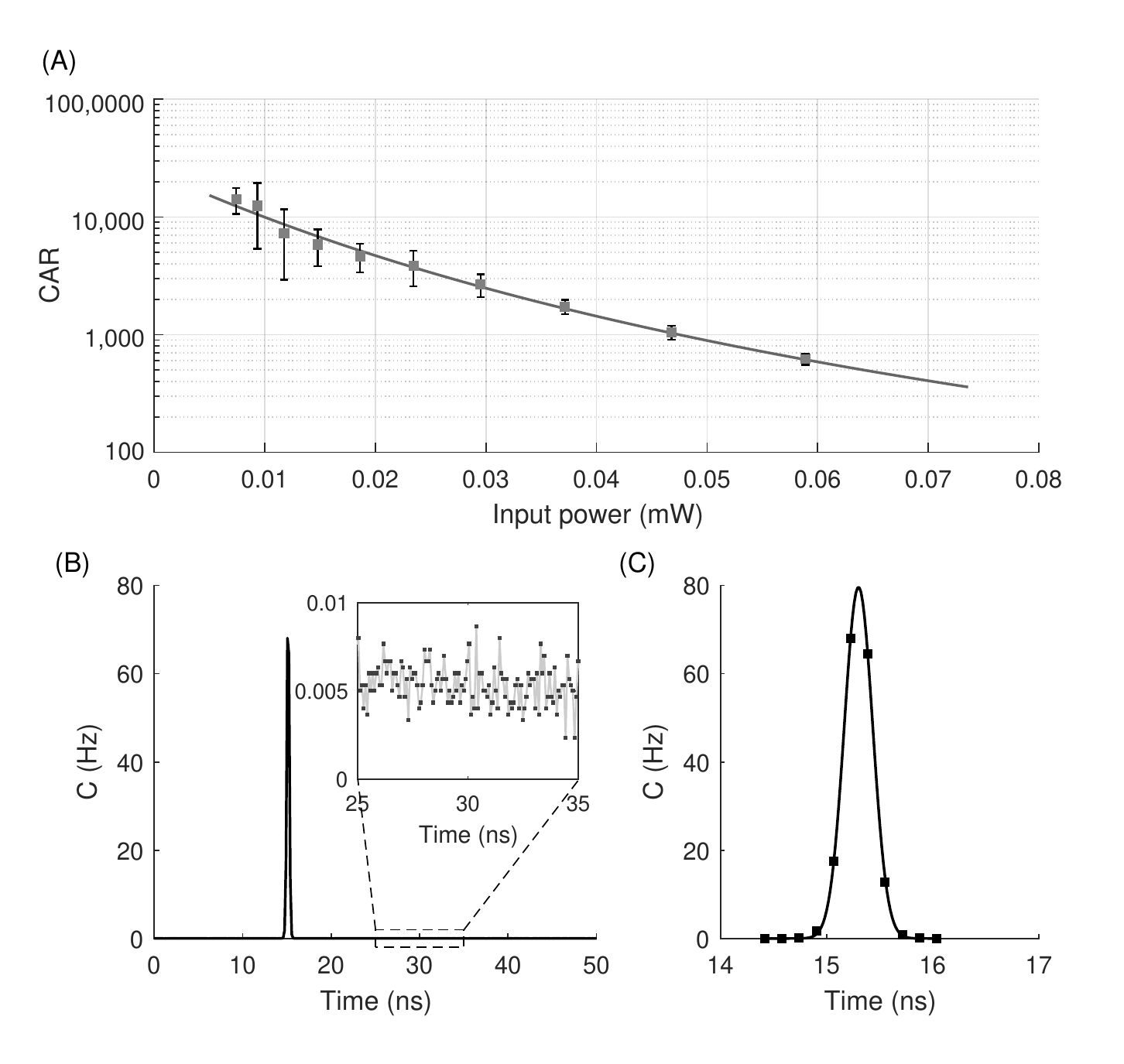}
\caption{Pair generation. (A) Coincidences-to-Accidentals Ratio (CAR) versus (cw) optical pump power in the feeder waveguide before the microring. The error bars are one standard deviation, calculated as described in the text. The highest measured CAR was $12,105\, \pm\, 1,821$. (B) The start-stop coincidence counting histogram for the highest CAR value. The inset shows a segment of the accidental coincidences. (C) Fit of the coincidence peak using a Gaussian function, with FWHM of 0.315~ns.} 
\label{fig-3}
\end{figure*}

\subsection{Coincidences-to-Accidentals Ratio (CAR)}
Fig.~3(A) shows the measurements of CAR versus input (cw) pump power. Raw two-fold coincidence counts ($C_{\text{raw}}$) and accidental coincidence counts ($A_{\text{raw}}$) between the generated photon pairs were measured for typical acquisition times of 30-300 seconds and binned into a histogram (one for each input pump power level). [The complete list of integration times is (from highest pump power to lowest power): 30, 30, 30, 120, 120, 180, 240, 600, 2100, and 3000 seconds.] The uncertainties in $A_{\text{raw}}$ are one standard deviation values of counts in bins away from the peak (start-stop delays were measured upto 100~ns time difference), and were propagated to generate the error bars in the CAR plot. Coincidences due to dark counts were measured separately, but since their contribution was negligible, they were not subtracted from the measurements. Each histogram peak was fitted by a Gaussian function, whose FWHM was typically 0.31~ns. The histogram of start-stop coincidences (measured bin counts divided by the measurement time in seconds) which resulted in the highest CAR is shown in Fig.~3(B), along with a segment of the accidental coincidences in the inset figure. As shown in Fig.~3(C), the peak was well fit by a Gaussian function. Its two-sided root-mean-square width defined the time window over which the (fitted) coincidence counts were integrated to yield $C$, and the same width was used to calculate the integrated averaged accidentals count, $A$, with CAR defined as $\text{CAR} = C/A$. (This number is less than $C_{\text{raw}}/A_{\text{raw}}$, since the accidentals counts are more or less flat over the integration window, whereas the coincidences are peaked.) 

The highest CAR was $12,105 \pm 1,821$ measured using an integration time of 3,000 seconds, when the pair generation brightness (defined as the pair generation rate divided by the FWHM of the loaded quality factor) was $8 \times 10^3\ \text{pairs}.\text{s}^{-1}.\text{GHz}^{-1}$. At the highest power values used here, CAR = $532 \pm 35$ with an integration time of 30 seconds, at a pair generation brightness of $550\times 10^3\ \text{pairs}.\text{s}^{-1}.\text{GHz}^{-1}$. Dividing further by the square of the optical pump power in the feeder waveguide (since the SFWM process is quadratic in the pump power) calculates the spectral brightness of the source, equal to $1.5 \times 10^8 \ \text{pairs}.\text{s}^{-1}.\text{GHz}^{-1}.\text{mW}^{-2}$ at the highest CAR value and $1.6 \times 10^8 \ \text{pairs}.\text{s}^{-1}.\text{GHz}^{-1}.\text{mW}^{-2}$ at the lowest CAR value (the two numbers are approximately the same, as is expected). The spectral brightness of this microring device is two orders of magnitude higher than reported by other groups~\cite{Mazeas:16}, a factor of twenty four higher than our previous results~\cite{Savanier:16}, and is a factor of two greater than that of the silicon microdisk\cite{Jiang:15}. 

CAR decreased at higher pump powers, as expected, and since the pair generation rate was higher, a shorter integration time was required to achieve reasonably small uncertainty error bars. CAR values in the tens of thousands have been measured in SPDC pair generation\cite{Harder:13}, but not yet shown in silicon photonics where values have been in the few thousands\cite{conf-Preble-2015,Lu:16}. (The CAR metric can be inflated by un-naturally narrowing the coincidence window, e.g.,~less than the timing jitter of the detectors~\cite{conf-Preble-2015}.) Achieving high CAR depends on low detector noise, suppression of pump and scattering noise, improvement of the stability of pair generation, and improving factors such as loss in the device and the experimental setup, which lead to broken pairs and increase the rate of accidental coincidences. 

\begin{figure*}[ht]
\includegraphics[width=5.8in, clip=true]{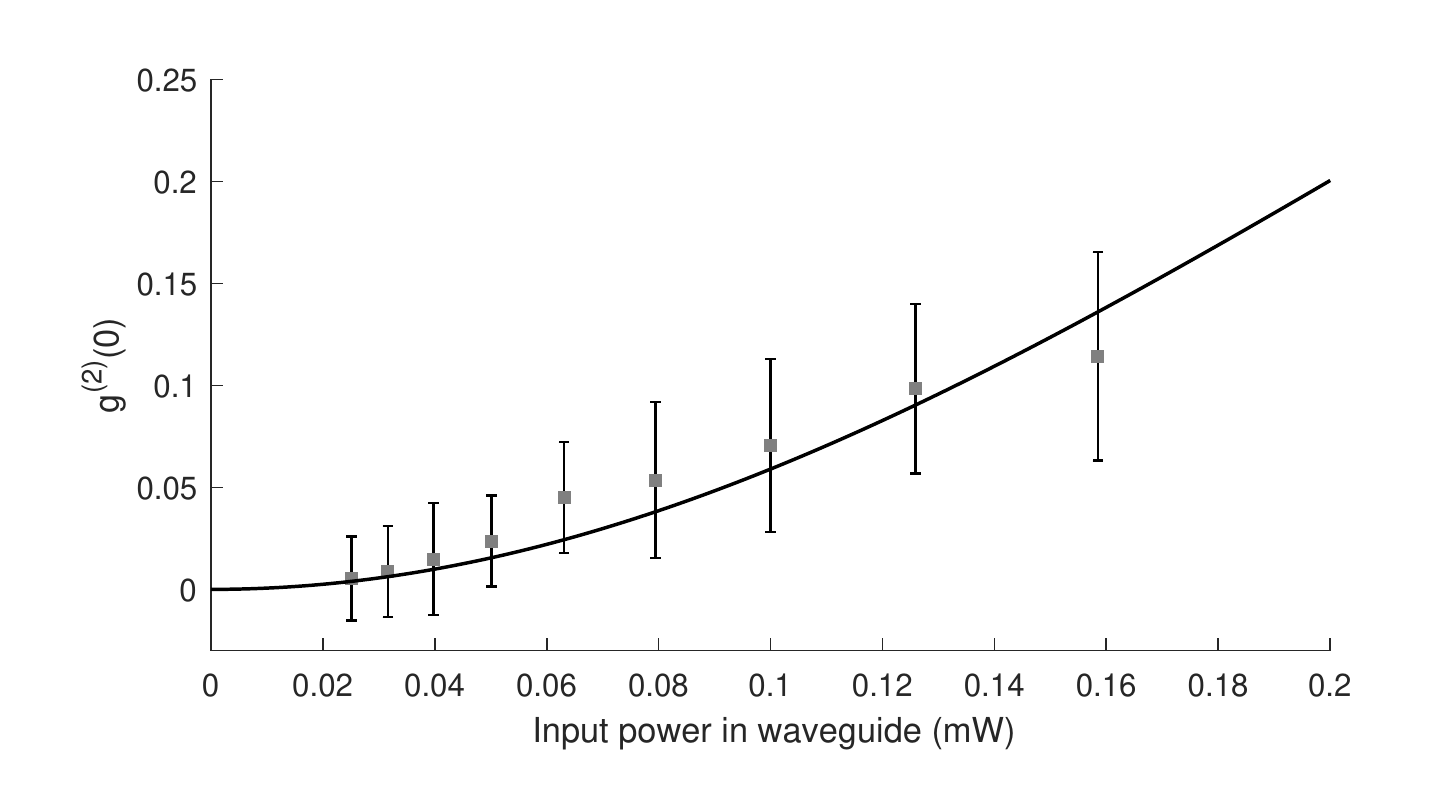}
\caption{Heralded single photon generation. Conditional self-correlation (heralded auto-correlation) $g^{(2)}_H(0)$ measured using the setup shown in Fig.~1(B). The error bars are one standard deviation. The lowest measured  $g^{(2)}_H(0)$ was $0.00533\, \pm\, 0.021$.}
\label{fig-4}
\end{figure*}

\subsection{Heralded single-photon generation}
Detecting one photon of the pair results in a heralded single-photon source, since the other photon is expected to show non-classical anti-bunching behavior. Fig.~4 shows the heralded (i.e.,~conditional) single-photon second-order self-correlation function, $g^{(2)}_H(0)$, obtained by detecting one of the generated photon pair as a herald, and measuring the self-correlation of the other photon in the presence of the herald. The normalized value of the photon correlation measurement on the heralded single photons at zero time delay was calculated using the formula~\cite{Beck2007} $g^{(2)}_H(0)=\frac{N_{ABC}N_{A}}{N_{AB}N_{AC}}$, where $N_{A}$ is the average photon detection rate on the heralding SNSPD detector [labels are shown in Fig.~1(B)], double coincidences $N_{AB}$ and $N_{AC}$ correspond to average rates of simultaneous events on one of the detectors (B or C) and the heralding SNSPD detector (A), and triple coincidences $N_{ABC}$ correspond to average rates of simultaneous events on all three detectors. The arrival times of events at the TDC module were synchronized by selecting appropriate lengths of BNC cables. Coincidences were defined as simultaneous detections within a 5~ns time window, measured directly by the TDC hardware (calculating coincidences between combinations of input channels without software post-selection). Counting times varied from 100 seconds for the higher pump powers to 600 seconds for the lowest pump power. The fitted line in Fig.~4 has the sigmoidal functional form, $g^{(2)}_H(0) = a P^2/(1+a P^2)$, where $a = 6.3\ \text{mW}^{-2}$ is a fitted coefficient. This fitting form is based on the fact that $g^{(2)}(0)$ is proportional to the biphoton rate\cite{PhysRevA.90.053825}, which in SFWM, is quadratic in the pump power, $P$, at low values, but saturates (to 1) as $P$ increases. Values as low as $g^{(2)}_H(0) = 0.0053 \pm 0.021$ were directly measured (the errorbar is one standard deviation uncertainty), for a measured heralding rate of $N_{A} = 18$~kHz. Even at the highest power values used in this sequence of measurements, $g^{(2)}_H(0) = 0.11 \pm 0.051$, well below the classical threshold, at a heralding rate of $N_{A} =340$~kHz. 

The heralding (Klyshko) efficiency, defined as $N_{AB}/(N_{A} \times D)$ where $D=0.65$ is the detection efficiency of the heralded photon, was calculated to be between 3\% and 4\% for the input powers shown in Fig.~4 (agreeing with the alternative calculation method, based on $N_{ABC}$ and $N_{BC}$, described in Ref.~\cite{Lu:16}). These are typical values for silicon photonic SFWM sources, and are about a factor of 10 inferior to those of SPDC based pair sources or glass integrated optics (at 700~nm wavelengths)\cite{Spring:13}. The main reason for the lower efficiency is loss: the sum of the fiber-to-waveguide loss and the insertion loss of the filters is about 7-10 dB for each of the photons in the current experimental configuration. Improving the Klyshko efficiency is a topic of current research, e.g., using active delay/multiplexing schemes to raise the heralding rates significantly\cite{Kaneda:15}. Another approach is to adjust the positioning between the micro-resonator (in the reported case\cite{Lu:16}, a microdisk resonator with an under-cut ``pedestal'') and the input waveguide (in the reported case, a tapered optical fiber). In our device, the position of the waveguide near the microring is fixed during lithography, and requires iterative fabrication to optimize. 

\subsection{Energy-time entanglement}
The generated photon pair is expected to demonstrate energy-time entanglement which can be investigated through a Franson-type two-photon interference experiment, by violating Bell's inequality~\cite{franson1989bell,kwiat1993high}. Such measurements have already been shown for several silicon photonic pair-generation devices~\cite{Harada2008,Grassani:15,Wakabayashi:15,Suo:15,KumarOpEx15,Mazeas:16}, and high values of visibility are confirmed for the microring device measured here. Fig.~5(B) shows the measurement of visibility fringes using a folded Franson interferometer configuration, in which both the signal and idler photons co-propagate through the same delay-line interferometer (DLI)~\cite{thew2002experimental}. Fig.~5(D) shows the measurements of the unfolded Franson interferometer configuration, in which two separate DLI's were used, one for the signal photons and one for the idler photons. Two fiber-coupled, polarization-maintaining, piezo-controlled DLI's, each with an FSR of 2.5~GHz and peak-to-valley extinction ratio approximately 25 dB were used in these measurements. Unlike in other experiments~\cite{KumarOpEx15,Mazeas:16}, no active DLI stabilization was required, considerably simplifying the experimental requirements. Approximately the same number of counts were measured here in 5~seconds as took nearly 50 times longer in other experiments~\cite{Mazeas:16}.

\begin{figure*}[ht]
\includegraphics[width=5.8in, clip=true]{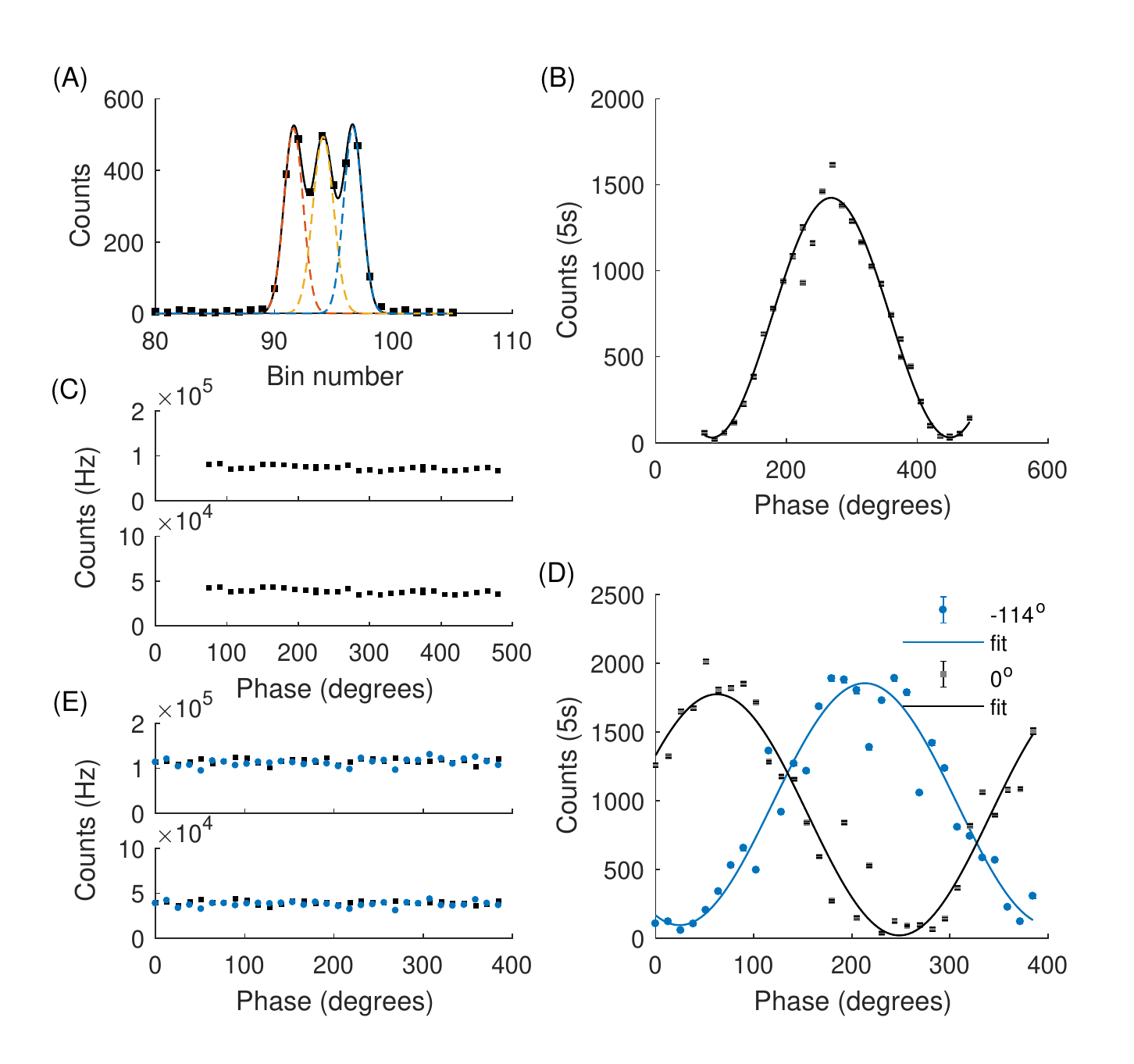}
\caption{(A) Representative histogram for the measurement of energy-time entanglement (at a particular phase setting of the DLI's), with an acquisition time of 5~s. The solid black line, showing the sum of the three Gaussians shown in red, blue and green dashed lines, fits the black dots which show the binned coincidence measurements. (B) Two-photon interference pattern measured using the folded Franson interferometer configuration. Grey dots (with errorbars): measured experimental data (coincidence counts), black line: fit. (C) The singles counts for the folded interferometer, measured at the same time as the two-photon coincidences. (D) Two-photon interference pattern measured using the un-folded Franson interferometer configuration. The interference pattern for two different phase settings on the second delay-line interferometer are shown. Grey and blue dots (with errorbars): experimental data, black solid and dashed lines: fit. (E) The singles counts for the unfolded interferometer. } 
\label{fig-5}
\end{figure*}

As shown in Fig.~5(A), the measured coincidences show three peaks, where the outer peaks correspond to the `short-long' (i.e., one of the photons goes through the short arm of its DLI and the other goes through the long arm of its DLI) and `long-short' coincidences, which do not change with phase tuning of the DLI's. The central peak consists of the interference of the `short-short' and `long-long' paths, whose peak amplitude varies sinusoidally with the phase difference between the two DLI's. The fitting uncertainty (one standard deviation) is shown as the errorbar in Figs.~5(B) and 5(D) and is too small to be visible. In the unfolded interferometer, the phase of one of the DLI's (i.e., the phase delay between the short arm and the long arm of that DLI) was held constant, and the phase of the other DLI was swept over approximately one free spectral range. The DLI's were tuned by voltage; as expected, the voltage required to tune the folded Franson interferometer, where both signal and idler photons experience the phase tuning, over a full period (3.86 V) was almost exactly one-half that of the unfolded Franson interferometer (7.82~V), where only the signal photon experiences the phase tuning. Although the signal and idler photons are at different wavelengths, separated by about 40~nm, the differential group delay accumulated over a few meters of fiber is negligible, compared to the timing jitter of the detectors.    

Proof of photon pair entanglement requires a two-photon interference pattern fringe visibility $V \geq 70.7\%$ (without necessarily providing a test of local realism). The fitted measurements showed $V$ clearly in excess of this threshold value, measured using a pump power in the feeder waveguide of about $50\ \mathrm{\mu m}$, resulting in a pair generation rate of about 68 kHz. From the raw data, $V_\text{data}^\text{(f)} = 98.9 \pm 0.6 \%$ for the folded Franson interferometer measurement, and $V_\text{data}^\text{(uf)} = 98.2 \pm 0.9 \%$ and $97.1 \pm 0.5 \%$ for the two phase settings of the unfolded configuration. From a fit to the measurements based on the non-linear least-square curve fitting algorithm in Matlab, $V_\text{fit}^\text{(f)} = 95.9 \pm 5.5 \%$ for the folded configuration, and $V_\text{fit}^\text{(uf)} = 97.8 \pm 14 \%$ and $90.3 \pm 14 \%$ for the unfolded configuration. In each case, the stated uncertainty is one standard deviation, but there are differences in the source of the uncertainty. For the data points, the uncertainty arises from the goodness-of-fit of the parameters of the Gaussian function used to fit the central peak [see Fig.~5(A)]. When the ensemble of points shown in Figs.~5(B) and 5(D) is fitted with a sinusoid function, the uncertainty then arises from the goodness-of-fit of those fitted parameters. These measurements confirmed the energy-time entanglement properties of the pairs, as shown by the sinusoidal variation of coincidences with phase, and in both cases, the flat singles rates (versus phase), shown in Figs.~5(C) and 5(E), show the absence of single-photon interference, as desired~\cite{kwiat1993high, franson1989bell}. 

\section{Discussion}
Recently, when using silicon photonic microring resonators, the reported maximum measured CAR values for pair generation (using equipment that is commercially readily available) have improved greatly from about 10 to over 10,000 and measured $g^{(2)}(0)$ values for heralded single-photon generation have decreased from about 0.2 to about 0.005, and Franson interferometry visibility has increased to about 98-99\%. The spectral brightness of our device, about $1.6 \times 10^8\ \text{pairs}.\text{s}^{-1}.\text{GHz}^{-1}.\text{mW}^{-2}$, was significantly higher than reported for other silicon microrings, and in fact, a factor of two higher than the microdisk resonator\cite{Jiang:15}. Measurements reported here took between a few seconds and a few minutes per data point.   

These experiments benefited from advances in single-photon detectors, and several other experimental improvements compared to our earlier experiments. Readout from an integrated monitoring photodiode across the microring helped align the resonance to the laser wavelength. The temperature of the microchip was accurately stabilized. The polarization state of the input light and fiber-waveguide alignment were actively monitored and controlled. We expect that with better packaging and higher levels of integration, performance may improve further, and devices may be tested for extended durations. The performance of the microring itself may still improve since the quality factor of the microrings is not at the achievable limit\cite{borselli2005beyond}. 
The fiber-resonator positioning challenges of the undercut microdisk structure~\cite{Lu:16} suggest that the microring device may be easier to package and use practically. On the other hand, the approach of Ref.~\cite{Lu:16} also presents an opportunity to optimize the resonator coupling coefficient for a particular application, e.g., improving heralding efficiency \textit{in situ\/}, which is not possible here. 

Taken together, the measurements reported here support and augment the growing evidence in favor of silicon (i.e., semiconductor) microring resonators as a viable architecture for optically-pumped photon pair and heralded single photon generation at telecommunications-compatible wavelengths, and establish a new level of performance of silicon-photonics devices, which (we hope) may now be considered seriously for potential implementation in experiments and deployed systems. The optical pump requirements are simple, and the pump could also be integrated into the silicon photonics manufacturing platform\cite{Wang:17}. Such devices may be inexpensive to manufacture (in volume) and could, for some applications, conveniently replace the crystal- or fiber-based instruments / assemblies used today both in the laboratory and in the field. 

\section*{Funding}
National Science Foundation (NSF) (ECCS 1201308, EFMA-1640968 ``ACQUIRE: Microchip Photonic Devices for Quantum Communication over Fiber''). 

\def\baselinestretch{1}\selectfont

\bibliographystyle{osajnl}
\bibliography{../../../../BibTeX_files/Quantum}

\end{document}